\documentclass[10pt,conference]{IEEEtran}
\IEEEoverridecommandlockouts
\usepackage{cite}
\usepackage{amsmath,amssymb,amsfonts}
\usepackage{graphicx}
\usepackage{textcomp}
\usepackage{xcolor}
\usepackage{algorithm}
\usepackage{algpseudocode}
\usepackage{hyperref}
\usepackage{graphicx}
\usepackage{subfigure}
\usepackage{multirow}
\usepackage{multicol}
\usepackage[breakable]{tcolorbox}
\usepackage{pifont}
\usepackage{enumitem}
\usepackage{amsmath}
\usepackage{wasysym}
\usepackage{bbm}

\usepackage{xspace}

\newcommand{\etc}{etc.}

\usepackage[numbers,sort&compress]{natbib}
\usepackage[normalem]{ulem}
\useunder{\uline}{\ul}{}
\usepackage{threeparttable}
\usepackage{makecell}
\usepackage{tcolorbox}
\usepackage{booktabs}

\usepackage{listings}

\def\BibTeX{{\rm B\kern-.05em{\sc i\kern-.025em b}\kern-.08em
    T\kern-.1667em\lower.7ex\hbox{E}\kern-.125emX}}

\usepackage{tikz}
\newcommand*{\circled}[1]{\lower.7ex\hbox{\tikz\draw (0pt, 0pt)%
    circle (.5em) node {\makebox[1em][c]{\small #1}};}}
\newcommand{\paragraphwithoutdot}[1]{\vskip 0.01in \noindent {\bf #1}}
\renewcommand{\paragraph}[1]{\vskip 0.01in \noindent {\bf #1.}}

\begin{document}
\title{ToolCoder: Teach Code Generation Models to use API search tools}
\author{\IEEEauthorblockN{Kechi Zhang}
\IEEEauthorblockA{
Key Lab of High Confidence Software \\
Technology, MoE (Peking University) \\
Beijing, China \\
zhangkechi@pku.edu.cn}
\and
\IEEEauthorblockN{Huangzhao Zhang}
\IEEEauthorblockA{
Key Lab of High Confidence Software \\
Technology, MoE (Peking University) \\
Beijing, China \\
zhang\_hz@pku.edu.cn}
\and
\IEEEauthorblockN{Ge Li* \thanks{* Corresponding authors}}
\IEEEauthorblockA{
Key Lab of High Confidence Software \\
Technology, MoE (Peking University) \\
Beijing, China \\
lige@pku.edu.cn}
\and
\IEEEauthorblockN{Jia Li \male}
\IEEEauthorblockA{
Key Lab of High Confidence Software \\
Technology, MoE (Peking University) \\
Beijing, China \\
lijia@stu.pku.edu.cn}
\and
\IEEEauthorblockN{Zhuo Li}
\IEEEauthorblockA{
Key Lab of High Confidence Software \\
Technology, MoE (Peking University) \\
Beijing, China \\
lizhmq@pku.edu.cn}
\and
\IEEEauthorblockN{Zhi Jin*}
\IEEEauthorblockA{
Key Lab of High Confidence Software \\
Technology, MoE (Peking University) \\
Beijing, China \\
zhijin@pku.edu.cn}
}

\maketitle
\begin{abstract}
Automatically generating source code from natural language descriptions has been a growing field of research in recent years. However, current large-scale code generation models often encounter difficulties when selecting appropriate APIs for specific contexts. These models may generate APIs that do not meet requirements or refer to non-existent APIs in third-party libraries, especially for lesser-known or private libraries. 
Inspired by the process of human developers using tools to search APIs, we propose ToolCoder, a novel approach that integrates API search tools with existing models to assist in code generation and API selection.
To teach our model to use tools, we introduce an automated data annotation method using ChatGPT to add tool usage information into the source code data and fine-tune code generation models. 
During inference, we integrate API search tools into the generation process so that our model can automatically use the search tool to get suggestions when selecting an API. 
Our experimental results demonstrate that ToolCoder exhibits excellent performance and generalization across five public and private library code generation benchmarks, with at least 6.21\% improvement on average pass@1 metrics and 9.64\% improvement on average pass@10 metrics compared to state-of-the-art methods.
Furthermore, we show that our relatively small ToolCoder model is comparable to one of the current best models, GPT-3.5, highlighting the potential of incorporating programming tools into the code generation process.
\end{abstract}

\section{Introduction}
\label{sec:intro}
Automated code generation has become increasingly important due to the significant effort required to manually write source code, especially for complex software. Deep learning techniques, particularly language models, have shown great promise in generating high-quality source code from natural language requirements. Currently, pre-trained code generation models are considered the state-of-the-art solution for various code generation tasks, such as CodeX \cite{humaneval},  ChatGPT \cite{gpt3, instructGPT} and CodeGen \cite{codegen} models. 

\begin{figure}[t]
\centering
  \includegraphics[width=0.55\columnwidth]{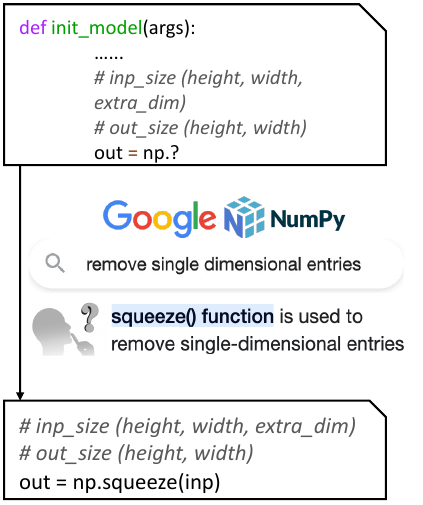}  
\caption{An illustrative example of the process of human programmers selecting the proper API during coding. Programmers summarize their demands into a query \textit{(remove single-dimensional entries)} and use the search engine tool or documentation search tool to get the proper API suggestion \textit{(np.squeeze)}.}
\label{fig:intro}
\end{figure}

Accurately selecting appropriate application programming interfaces (APIs) is essential for pre-trained models to generate code. API selection is crucial for accurately expressing program semantics and efficiently addressing problems.
However, there are too many existing third-party libraries and their APIs, and new APIs are constantly being developed. Existing models often find it challenging to select APIs accurately and will generate non-existent APIs or APIs that do not meet requirements.
For example, according to our preliminary experiments on NumpyEval and PandasEval \cite{codegenapi}, a popular code generation model CodeGen-2B generates more than 26\% code that contains an incorrect API. Furthermore, for security and functionality reasons, industrial companies often build private libraries for internal use only. For these private libraries that are not publicly available, the error rate will increase to more than 90\%. These third-party public libraries or private libraries provide so many APIs that code generation models have never seen, resulting in the model being unable to generate API-oriented code.
Therefore, it is worth exploring methods to improve code generation models to generate accurate source codes using these domain-specific or private library APIs.

To assist code generation models in selecting appropriate APIs during the generation process, we draw inspiration from human programmers' perspectives. 
In most programming scenarios, programmers can use search tools to get suggestions from web sources or library documents when selecting an API.
Figure \ref{fig:intro} shows an example of human programmers selecting the proper API during coding. When encountering an API usage scenario, programmers summarize their needs into a query and use existing API search tools to search for suitable APIs, such as Google search engines or documentation search tools for specific libraries. Then, according to the search results, programmers can choose the proper API. This programming process using tools to retrieve and determine API usage improves programming efficiency, improves accuracy, and reduces the risk of errors. 
It motivates us to investigate approaches to teach code generation models to use search tools to find suitable APIs.

In this paper, we propose ToolCoder, a low-cost and efficient solution that integrates API search tools into pre-trained code generation models, mimicking how programmers solve this problem. 
To help models learn to use tools, we propose an automated data annotation method with ChatGPT to add tool usage information into the source code data and use the annotated dataset to fine-tune code generation models.
Specifically, our approach utilizes the in-context learning ability of large models to annotate a special tool-augment dataset at a low cost. We employ parameter-efficient fine-tuning to improve the training efficiency.
During inference, we integrate API search tools into the decoding process of our model, allowing the model to learn to use external tools autonomously.

We extensively evaluate our proposed ToolCoder with pass rate metrics \cite{humaneval}.
\ding{182} We evaluate ToolCoder on three public library benchmarks. 
Our model achieves significant improvements over state-of-the-art baselines with at least 10.11\%, 3.26\%, and 1.39\% pass@1. Our relatively small ToolCoder is even comparable to one of the current best language models, GPT-3.5. 
\ding{183} We further evaluate our model on two private library benchmarks. 
By switching to the appropriate search tool, our ToolCoder can be easily transferred to these private library scenarios and achieves stable improvement. 
Our model exhibits better generalization performance and raises at least 6.21\% improvement on the average pass@1 metrics for all five benchmarks.
\ding{184} We also conduct an ablation study to analyze the different settings in our experiments, including the dataset, training, and inference settings. Results prove the effectiveness of different designs in our approach.

Our contributions in this paper can be summarized as follows:
\begin{itemize}
    \item To the best of our knowledge, we are the first to incorporate a programming tool into code generation models. Our results highlight the importance of models' ability to use tools. 
    \item We propose an automatic method for annotating datasets in software engineering. This low-cost and efficient annotation framework uses powerful ChatGPT to annotate API datasets with public source code datasets, reducing the manual effort required to create private annotated datasets. Our dataset construction method can also be easily transferred to other tasks.
    \item We propose ToolCoder, which incorporates the ability to use API search tools into pre-trained code generation models and improves the performance on API-related code generation tasks. Our approach outperforms existing API-oriented baselines on multiple popular API-related code generation benchmarks.
\end{itemize}

\section{Motivating Examples}
\label{sec:motivation}

In this section, we examine the limitations of the current code generation models when selecting a suitable API and how existing search tools can aid in API selection. By exploring these issues, we hope to provide context for our research and explain our motivation for proposing a new approach to address these challenges.

\subsection{Limitations of current code generation models in selecting suitable APIs}
\label{sec:motivating_codegen}
\begin{figure}[t]
\centering
  \includegraphics[width=0.7\columnwidth]{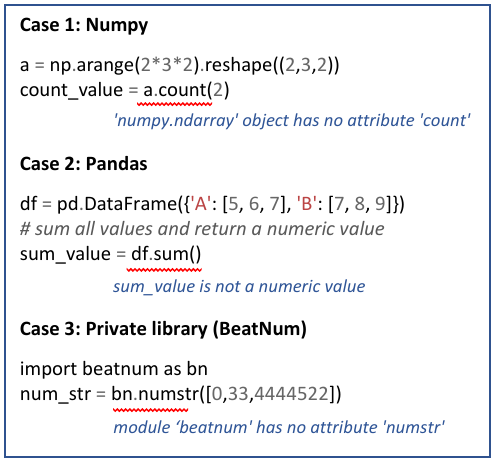}  
\caption{Failure Cases of \textit{CodeGen-2B} model in selecting APIs, including generating non-existing APIs on public libraries (\textit{Case 1}), generating unqualified APIs (\textit{Case 2}), and lack of API-related knowledge on private libraries (\textit{Case 3}).}
\label{fig:motivating_codegen}
\end{figure}
Application Programming Interfaces (APIs) are essential to modern software development. APIs allow developers to integrate pre-existing code and services into their applications. Using APIs can significantly reduce the time and effort required to develop complex software systems.

However, selecting the proper API remains challenging for code generation models.
Due to the proliferation of third-party libraries and their associated APIs, existing code generation models often need help choosing APIs. 
Here we choose the popular model CodeGen-2B to test its performance on API selection. Figure \ref{fig:motivating_codegen} shows three failure examples of the generated code for selecting APIs. 
\ding{182} \textbf{Case 1:} The CodeGen-2B model has the potential risk of generating APIs that do not exist, even on such a popular and common third-party library as NumPy. There is no \textit{count} API in the Numpy library, but CodeGen2B still generates. 
\ding{183} \textbf{Case 2:} It may also use the wrong API and generate unqualified code. \textit{df.sum()} will return with a Pandas Series type but not the required numeric value. It shows that these existing code generation models still have challenges choosing an appropriate API to implement a given requirement. 
We conduct a statistical experiment to analyze the generated code on two benchmarks, NumpyEval and PandasEval \cite{cert}, and find that more than 26\% of the APIs generated have the problems mentioned above. 
We also conduct experiments on private library benchmarks such as BeatNumEval \cite{codegenapi}. Private libraries are widespread in real code scenarios, and they are not public for security and functional reasons. Experiments show that for those APIs in private libraries, the failure rate will increase to more than 90\%.
\ding{184} \textbf{Case 3:} We find the CodeGen-2B lacks corresponding private knowledge and concocts an incomprehensible API for the private library BeatNum. It shows that existing code generation models have limitations in API generation and are not specifically optimized for using APIs. In this paper, we aim to address these API selection issues.

\subsection{Existing search tools to aid API selection}
\label{sec:motivating_tool}

Drawing inspiration from the approach programmers employ in selecting an API, our study reveals that existing search tools can provide API recommendations. For example, in Figure \ref{fig:intro}, the developer needs to use the \textit{numpy} library to remove the extra single dimension in \textit{input\_size}. The developer turns to \textbf{online search engine tools} or \textbf{library documentation search tools} and gets the proper API suggestion \textit{np.squeeze}. These two search tools can play a significant role in selecting APIs. A comparison of these two types of search tools is given in Table \ref{tab:motivating_tools}. We will analyze the use of these two types of search tools. 
\begin{table}[t]
\setlength\tabcolsep{5pt}
  \centering
  \caption{Comparisons of two types of search tools for API selection.}

\begin{tabular}{lll}
\toprule
                    & Online Search Engine                                                                 & Documentation Search                                                                           \\ \midrule
\begin{tabular}[c]{@{}l@{}}Knowledge \\ Resources\end{tabular} & \begin{tabular}[c]{@{}l@{}}Programming Community \\ or Tutorial Websites \\ {\scriptsize (StackOverFlow, datagy.io, \etc)}\end{tabular} & Library Documentation                                                                          \\ \midrule
API Type            & \begin{tabular}[c]{@{}l@{}}Public libraries,\\ especially those well-known \\ and widely-discussed\end{tabular}   & \begin{tabular}[c]{@{}l@{}}Any APIs,\\ including public and \\ private libraries\end{tabular} \\ \midrule
Advantages           &  \begin{tabular}[c]{@{}l@{}}Practical and Accurate \\ Rich sources \\ Keep updating\end{tabular}                                                                                   &     \begin{tabular}[c]{@{}l@{}}Wide coverage \\ Detailed explanation \\ Stable \end{tabular}                                                                                     \\ \midrule
Example Tools        & \begin{tabular}[c]{@{}l@{}}Google, Bing, \\ DuckDuckGo  \end{tabular}                                                        &\begin{tabular}[c]{@{}l@{}}NumPydoc, \\ Pandasdoc, \\ Private documentations  \end{tabular}                                                                    \\ 
\bottomrule
\end{tabular}
  \label{tab:motivating_tools}
\end{table}

\subsubsection{Online Search Engine Tool}

Online search engine tools provide rich information on various API usage scenarios. Human programmers share their experience in solving various programming problems on various community and tutorial websites such as \textit{StackOverFlow}\footnote{https://stackoverflow.com/} and \textit{datagy.io}\footnote{https://datagy.io/}. They organize and summarize the API suggestions used for different problems. Formally, these online API suggestions are usually displayed in the form of programming experience sharing or question and answer.
When other people encounter similar problems, search engines can use this information well, and programmers only need to provide a question query. These search engines regard these community websites as knowledge resources and can provide helpful API suggestions, especially for those public libraries that are well-known and widely discussed. 
Since these online programming experiences on the website are constantly updated and provide a variety of practical scenarios, we can often get more accurate API suggestions from these online search engine tools. These mature commercial search engine tools such as \textit{Google}\footnote{https://www.google.com/}, \textit{DuckDuckGo}\footnote{https://duckduckgo.com/} can provide accurate, instant, and fast search responses and are widely used by human programmers. 

\subsubsection{Documentation Search Tool}

Since lesser-known public libraries or private libraries have few discussions on the online community websites, human programmers also turn to library documentations for API suggestions.
Documentations are usually available for public libraries and private libraries. Therefore, it can provide rich information for any API usage scenario.
Documentations provide detailed explanations for each API, with a broad convergence over the corresponding library. The documentation contains detailed and accurate explanations of the parameters and usage of the API. It is the most detailed place to learn how to use an API. Since the documentation does not change frequently, its results are more stable. 
Formally, API information in the documentation is usually given in pairs of API and corresponding comments. We can use \textit{BM25} \cite{BM25} or other semantic similarity scores as search metrics to search for comments that meet the requirements and find the corresponding API as the final suggestion for coding.

These various search tools are helpful for programming and selecting an API. Inspired by the API search process of human developers, we aim to incorporate these two types of search tools into code generation models. By letting the code generation model learn to use these online search engine tools or documentation search tools, our models can effectively navigate the vast amount of information available to identify the most relevant APIs. This approach enables our models to more accurately match APIs with specific needs.

\section{API Search Tool}
\label{sec:background}
\label{sec:background_tool}
To better present the methodology of our model, we first provide a brief introduction to the API search tool in this section. The proposed API search tool is an abstraction of the existing search tools for API selection and will be used as an external tool for our ToolCoder.

Following the motivating examples in Section \ref{sec:motivating_tool}, we develop the API search tool for code generation models based on these two categories of sources. \ding{182} For those commonly used public libraries such as \textit{numpy} and \textit{pandas}, we use \textbf{\textit{DuckDuckgo}} as the search engine because it provides a more convenient and automated method compared to other search engines. We use the search engine to search the relative content from several online community websites and extract the mentioned APIs with string regex matching. Since these contents have a richer introduction to the API, more accurate API suggestions can be obtained from the search engine. \ding{183} For lesser-known APIs or those in private libraries, we employ the \textbf{\textit{BM25}} score as our retrieval metric to search from the corresponding API documentation.

We then abstract the two types of search tools into a unified form: we use the notation $APISearch(query)\rightarrow answer$ to represent the call of API search tool, where $APISearch$ is the function name that abstracts different API search sources, $query$ denotes the search query and $answer$ indicates the return answer from API search tools that can be referred to for further code generation. In subsequent experiments, we serialize API search tool calls for model input. To differentiate from regular text, we surround the tool calls with special tokens by starting with $\left \langle API \right \rangle $ and ending with $\left \langle /API \right \rangle $. Examples can be viewed in Figure \ref{fig:pipeline}. We set $\left \langle API \right \rangle $, $\left \langle /API \right \rangle $, and $\rightarrow$ as special tokens in our model vocabulary.

\section{ToolCoder}
\label{sec:model}
\begin{figure*}[t]
\centering
  \includegraphics[width=1.6\columnwidth]{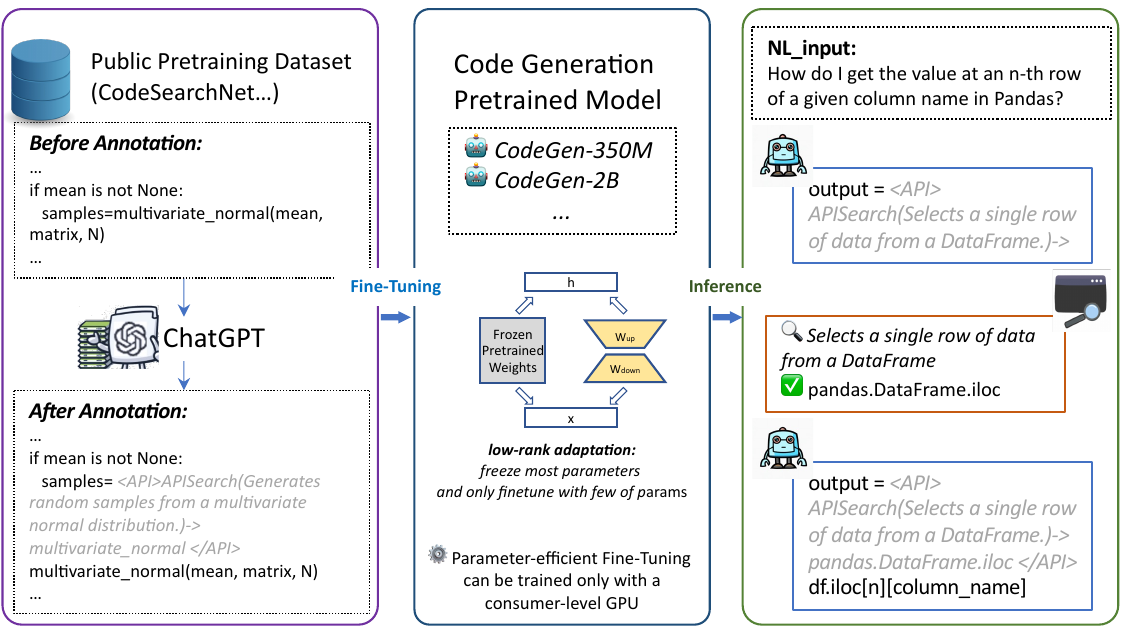}  
\caption{The pipeline of our approach ToolCoder. The pipeline has three main parts: (1) Automatically Annotate Tool-augmented Dataset with ChatGPT, (2) Parameter-efficient Fine-tune existing pre-trained code generation model with the annotated dataset, and (3) Inference of the fine-tuned model enhanced with API search tools.}
\label{fig:pipeline}
\end{figure*}

In this section, we present our approach ToolCoder for selecting and using APIs in coding practices. The goal of our approach is to train a model that can effectively select and use appropriate APIs based on existing partial code. To achieve this goal, we decompose our approach into three modules, including data annotation, fine-tuning, and inference. The three modules work in a pipeline as shown in Figure  \ref{fig:pipeline}.
We will describe the details in the following subsections.

\subsection{Automatic Data Annotation}
In order for the model to learn to use the API search tool, we first need a dataset that includes the source code and associated tool call processes. As mentioned in Section \ref{sec:background_tool}, we abstract the search call process with the notation $\left \langle API \right \rangle  APISearch(query) \rightarrow answer \left \langle /API \right \rangle $. However, such datasets are not readily available. To address this issue, we propose automatically augmenting an existing source code dataset with the tool call notation using ChatGPT (gpt-3.5-turbo)\footnote{https://openai.com/}, which has demonstrated excellent few-shot and even zero-shot learning ability in many different language learning tasks already. This low-cost and efficient annotation method reduces the manual effort required to create private annotated datasets. Our data annotation process can be divided into three parts: \ding{182} base dataset selection, \ding{183} prompt selection, and \ding{184} filter and clean.

\paragraph{Base Dataset Selection}
For the base dataset, we choose to use the popular pre-trained dataset \textit{CodeSearchNet-Python} \cite{CSN} as the base dataset. It is a real-world programming dataset obtained from GitHub without any additional annotations. This dataset is already commonly used by many pre-trained code generation models, so we can assure as much as possible that our subsequent training will not affect the model's generalization performance on language generation and modeling ability. We use a simple length filtering method and randomly choose nearly 60k function-level source code from this dataset as the base dataset for our annotation method.

\begin{figure}[t]
\centering
  \includegraphics[width=0.8\columnwidth]{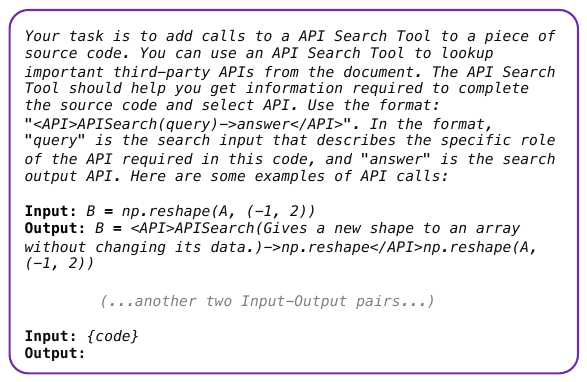}  
\caption{An exemplary prompt used to generate API-augmented datasets for the API search tool. In our setting, We selected a total of three human-written input-output pairs as part of the prompt, using three libraries: \textit{numpy}, \textit{pandas}, and \textit{matplotlib}.}
\label{fig:prompt}
\end{figure}

\paragraph{Prompt Selection}
Similar to \cite{toolformer}, to help generate the annotated dataset, we need to provide a detailed instruction for ChatGPT to specify its system role as a data annotator, shown in Figure \ref{fig:prompt}. To facilitate the quality of the generated datasets, we manually write three human-written input-output pairs as part of the prompt with three libraries including \textit{numpy}, \textit{pandas}, and \textit{matplotlib}. We choose these three libraries as the examples in the prompt because we are skilled in them, and they are also commonly used in the base dataset.
Based on our selected prompt and base dataset, we will ask the ChatGPT to annotate the tool-augmented dataset. We generate one annotated data for each base sample. The automatic annotation process lasted for four days.

\begin{table}[t]
\setlength\tabcolsep{5pt}
  \centering
  \caption{Statistics of the annotation dataset. }

    \begin{tabular}{l r}
        \toprule
        \textbf{Statistic} &  \\ 
        \midrule
        Dataset Size & 53,000 \\
        Avg. Annotation API & 3.2 \\
        Avg. Length (in words) before annotation & 186.24 \\
        Avg. Length (in words) after annotation & 211.49 \\
        Proportion of some third-party libraries & \begin{tabular}{@{}lr@{}}
            \textit{NumPy} & 24\% \\
            \textit{Pandas} & 13\% \\
            \textit{TorchData} & 0\%
        \end{tabular} \\
        \bottomrule
    \end{tabular}
  \label{tab:dataset_statistic}
\end{table}

\paragraph{Filter and Clean}
After getting all the generated results from chatGPT, we performed a series of simple filtering operations on the results to remove those abnormal data samples. We filter out nested API Search calls, control the number of API Search calls in a sample of less than 5, and ensure that at least one is an API call from a public library. We filter out those samples that are different from the source code after removing the API Search call. Furthermore, for the generated API answer in the search call, we check whether it is followed by the corresponding API in the generated code to ensure that the API search call is closely related to the specific code implementation. Finally, we cleaned and obtained the final data set of 53k, which will be used for subsequent fine-tuning. Table \ref{tab:dataset_statistic} shows the statistics of the final annotated dataset. We also count the proportion of some third-party library APIs in the dataset for reference in subsequent evaluation experiments. In the left part of Figure \ref{fig:pipeline}, we also give an example sample of the final dataset.

\subsection{Parameter-efficient Fine-tuning}
We leverage the annotated dataset to fine-tune a pre-trained language model to teach the model to generate the search tool call itself. 
To address the challenge of limited computational resources and improve the training efficiency, we propose restricting the number of meta-trainable parameters and layers in the pre-trained model and adopting a parameter-efficient fine-tuning approach that can efficiently adapt pre-trained models to new task types. In particular, we apply LoRA \cite{lora} to reduce trainable parameters.

Low-Rank Adaptation (LoRA) is a low-dimensional representation-based parameter-efficient tuning method. It injects trainable low-rank matrices into transformer layers to approximate the weight updates. 
For a pre-trained weight matrix $ W \in \mathbb{R}^{d \times k}$ , LoRA represents its update with a low-rank decomposition $W +\delta W = W +W_{down}W_{up}$ , where $W_{down} \in \mathbb{R}^{d \times r}  , W_{up} \in \mathbb{R}^{r \times k}$ are tunable parameters. LoRA generally applies this update to the attention linear projection matrices in the multi-head attention sub-layer in Transformer. For a specific input $x$ to the linear projection in multi-head attention, LoRA modifies the projection output $h$ as:

\begin{equation}
    h \leftarrow  h + s \cdot xW_{down}W_{up} , 
\end{equation}
where $s \ge 1$ is a tunable scalar hyperparameter. The illustration of LoRA is shown in the middle part of Figure \ref{fig:pipeline}.

In our training setting, we freeze most of the parameters in the pre-trained model and only apply LoRA on the query and value projections in the attention module for each transformer layer. As a result, we only need to train 0.18\% parameters in \textit{CodeGen-350M} and 0.09\% for \textit{CodeGen-2B}. It makes it possible to efficiently fine-tune models on a consumer-level GPU, such as Nvidia GeForce RTX 2080 (11GB RAM). 
The parameter-efficient tuning strategy significantly reduces the training computational burden in our experiments. It can achieve results comparable to full-parameter training with less computational resources and time. We will give a detailed analysis of the ablation experiment in Section \ref{sec:ablation}.

\subsection{Inference enhanced with Tools}
After training with the annotation dataset, the model can generate the API search calls during the code generation process. The pseudo-code description of the decoding process with API search tool procedure is in Algorithm \ref{alg:infer}.

\begin{algorithm}[h]
\scriptsize
\caption{Inference with API Search Tool}
\label{alg:infer}
\begin{algorithmic}[1]
\Procedure{InferWithTool}{model, input\_nl, maxlen}
\State Pass $input\_nl$ to the model and get predicted $token$
\State $output \gets [token]$
\State $i \gets 0$
\While{$i < maxlen$}
    \State $token \gets$ the last token of $output$
    \If{$token = \left \langle API \right \rangle $}
        \State $query \gets$ the following generated tokens between \textit{APISearch(}  and \textit{)$\rightarrow$}
        \State $response \gets$ Call API search tool with $query$
        \State Append \textit{$\left \langle API \right \rangle $APISearch(query)$\rightarrow$response$\left \langle /API \right \rangle $} to $output$
        \State $i \gets i + $ length of the call process
    \Else
        \State Pass $token$ to the model and get predicted token
        \State Append predicted token to $output$
        \State $i \gets i + 1$
    \EndIf
\EndWhile
\State \textbf{return} $output$
\EndProcedure
\end{algorithmic}
\end{algorithm}

During inference, we perform regular decoding until the model produces the $\left \langle API \right \rangle $ token, indicating that it next expects the response for an API call. At this point, we continue the decoding process and record the following generated tokens to get the query between $APISearch($ and $)\rightarrow$. Then we interrupt the decoding process and call the API search tool to get a response, and continue the decoding process after inserting both the response and the $\left \langle /API \right \rangle $ token. 

As mentioned in Section \ref{sec:background_tool}, we adopt different API search sources for different types of API usage. For those commonly used public libraries, we use the \textit{DuckDuckGo}, a popular online website search engine, to adopt content in-site search in several selected websites. For those lesser-known or private library APIs, there is no relevant online information. So we employ the BM25 score as our retrieval metric to search from the corresponding API documents. We encapsulate these search interfaces so that our ToolCoder can call search tools with high performance. 
In our experiment, we control the search delay within \textit{0.6s} to ensure high efficiency during the code generation process.

After the entire inference process is over, we use the regular matching method to remove the API search part from the generated code, that is, the part between $\left \langle API \right \rangle $ and $\left \langle /API \right \rangle $ to get the generated code. By using API search tools in this way, we can effectively address the challenge of selecting appropriate APIs and reduce the time and effort required for developers to find suitable APIs.

\section{Experimental Setup}
\label{sec:setup}
To assess the effectiveness of our approach, we perform a large-scale study to answer four research questions. In this section, we describe the details of our study, including datasets, metrics, and baselines.
\subsection{Research Question}
Our study aims to answer four research questions. 
In RQ1, we compare our ToolCoder to SOTA code generation models on three public library benchmarks. 
In RQ2, we conduct experiments on two private library benchmarks to show the generalization of our proposed model on those private libraries.
In RQ3, we conduct an ablation study to prove the contributions of different modules. 
In RQ4, we conduct a series of quality measures on the generated results and analyze the effectiveness and limitations of our method through detailed case studies.

\textbf{RQ1. How does ToolCoder perform compared to SOTA baselines on public library code generation?}
To evaluate ToolCoder's performance on public library code generation, we conduct experiments on three public library code generation benchmarks, including \textit{numpy}, \textit{pandas}, and \textit{torchdata}. We compare ToolCoder's performance with existing SOTA code generation baselines.

\textbf{RQ2. How does ToolCoder perform on private library code generation?} 
We select two private library benchmarks where the pre-trained language models have never encountered any private library APIs, and there is no relevant information available online. We evaluate ToolCoder's performance on these private libraries to demonstrate its generalization and versatility.

\textbf{RQ3. What are the contributions of different modules in our approach?}
Our approach pipeline consists of three modules: data annotation, fine-tuning, and inference. To analyze the effectiveness of our approach, we conduct an ablation study by varying settings in our pipeline, including the dataset, training, and inference search settings.

\textbf{RQ4. How is the quality of our generated code with ToolCoder?}
We evaluate the quality of generated code using ToolCoder by performing a case study analysis. Additionally, we analyze the effectiveness of our method and explain why our model works.

\subsection{Datasets}
Our experiments are conducted on three public library benchmarks, PandasEval, NumpyEval, and TorchDataEval, and two private library benchmarks, including MonkeyEval and BeatNumEval.
We choose these benchmarks to ensure our proposed method can be used in various API selection scenarios.

\subsubsection{Public library benchmarks}
\textbf{PandasEval} \cite{cert} is a domain-specific method or block generation benchmark for the Pandas library in Python.
PandasEval contains 101 test examples.
Each example corresponds to a programming problem of Pandas, containing the context code, the target method body (or block), and multiple test cases.
\textbf{NumpyEval} \cite{cert} is almost the same as PandasEval, apart from the domain. NumpyEval specifically targets the Numpy library in Python.
The benchmark also contains 101 test examples.
\textbf{TorchDataEval} \cite{codegenapi} is based on the TorchData library in Python.
TorchData is a newly released library, which is more likely to be unseen to the pre-trained models.
Therefore, this benchmark is proposed to evaluate the model against the unseen library containing 50 test examples.
In our experiments, our annotated dataset does not contain API code related to TorchData as shown in Table \ref{tab:dataset_statistic}, and our base pre-trained model does not contain these data during the pre-training phase, so this benchmark can also be used to demonstrate the generalization ability of our method on those APIs that are public but never seen by the code generation model.

\subsubsection{Private library benchmarks}

\textbf{MonkeyEval} \cite{codegenapi}, modified from PandasEval, is designed to evaluate the method generation model against the unseen library.
The Monkey library is crafted by modifying all Pandas-related keywords. \emph{e.g.}, ``pandas'' is converted to ``monkey'', ``dataframe'' is converted to ``knowledgeframe'', \etc.
The library construction process ensures that no information about the API names of these libraries is leaked in the online materials or any training datasets.
MonkeyEval converts all examples in PandasEval, leading to 101 test examples.
\textbf{BeatNumEval} \cite{codegenapi} is modified from NumpyEval, in the same way as PandasEval to MonkeyEval. BeatNumEval also has 101 test examples.
The pre-trained model has not seen the API in MonkeyEval and BeatNumEval, and the online search resources cannot provide any API-related information. So the API selection on these benchmarks will only rely on the API search tool we built on the documentation of these private libraries.

\subsection{Metrics}
Following the previous work, we use the metric pass rate \textbf{\textit{pass@k}} \cite{humaneval} for performance evaluation and take advantage of the provided unit tests to determine the functional correctness of code solutions. For each problem, we submit k code solutions for evaluation. If any of the k code solutions passes all ground truth test cases, the problem is considered solved. Then pass@k is the percentage of solved problems. In our experiments, we set $k = \{1,10\}$.

\subsection{Baselines}
We select six series of recent code generation models as baselines, including one of the most powerful models, GPT-3.5. These models can be divided into two categories: general models and API-oriented models.
\subsubsection{General Models}
\textbf{CodeT5} \cite{CodeT5} is an encoder-decoder pre-trained model for code-related tasks. It uses the identifier-aware pre-training task and has achieved SOTA results on many general code generation benchmarks. We use CodeT5-base with 220M parameters in our experiments.
\textbf{PyCodeGPT} \cite{cert} is a decoder-only pre-trained code generation model with 110M parameters. It is initialized with the GPT-Neo and is continually pre-trained with a large-scale code corpus in Python.
\textbf{CodeGen} \cite{codegen} is a series of decoder-only pre-trained code generation models with parameters varying from 350M to 16B. It casts code generation as a multi-turn conversation between a user and a system. CodeGen has shown strong ability on a variety of complex code generation tasks. Due to computational limitations, we use 350M and 2B versions in our experiments.
\textbf{GPT-3.5} \cite{gpt3, instructGPT} is one of the most powerful generation models from OpenAI. We use the ``\textit{gpt-3.5-turbo}`` model as it is the most cost-effective and performant model in the GPT3.5 family. As OpenAI states, it can be complemented with flexible natural language and programming language capabilities\footnote{https://platform.openai.com/docs/models/gpt-3-5}.

\subsubsection{API-oriented models}

\textbf{CERT} \cite{cert} is a generation approach designed for API-related code. CERT contains two modules: the sketcher and generator, each of which is fine-tuned independently with PyCodeGPT. It first predicts a sketch based on the NL description and generates the complete code based on the sketch. For each library, CERT requires a specially trained weight for generation. We use the released weight as two independent models: \textit{CERT-numpy}, \textit{CERT-pandas}.
\textbf{CodeGenAPI} \cite{codegenapi} is another API-oriented code generation model. It uses a two-stage pipeline to generate code: given an NL description, CodeGenAPI firstly uses a retriever model initialized with BERT \cite{bert} to find APIs from documents. Then it uses a generator initialized with CodeGen-350M to generate the complete code based on the retrieved API and problem description. We use the three released settings in their paper: \textit{CodeGenAPI}, \textit{CodeGen-retrieval}, and \textit{CodeGenAPI-retrieval}. The first setting only uses the trained generator without retrieval, and the latter two use the best-performing top2 retrieval results to assist generation.

\subsection{Implementation Details}
\label{sec:imple_details}
\paragraph{Training} Our model is implemented in the Pytorch framework, and we perform all the experiments on four RTX 2080-11GB GPUs. We initialize our ToolCoder by leveraging pre-trained weights of CodeGen-350M and CodeGen-2B. The training batch size is set to 8, and the total training epoch is set to 10. We use validation loss to determine the best checkpoint as the final model. 
\paragraph{Tool} \label{sec:imple_tool} When implementing the API search tool, we adopt in-site online search in \textit{datagy.io} as well as \textit{NumPy}\footnote{https://numpy.org/doc/}, \textit{Pandas}\footnote{https://pandas.pydata.org/docs/} and \textit{TorchData websites}\footnote{https://pytorch.org/data/} using the \textit{DuckDuckGo} for public library benchmarks. For private library benchmarks, we use provided \textit{Monkey} and \textit{BeatNum} library documentations to design an API search tool based on the BM25 algorithm. The tool's response for inference is considered as the first retrieved API. 
\paragraph{Inference} During the model generation process, we use temperature sampling with $T = 0.8$ and limit the sample budget to 10. Each experiment is run three times with random seeds and then averaged for the final results.

\section{Results and Analyses}
\label{sec:results}
\subsection{RQ1: Results for Public library API Code Generation}
\label{sec:public_rst}
\begin{table}[t]
\scriptsize
    \setlength\tabcolsep{4pt}
    \centering
    \caption{Pass rate of models on Public library benchmarks}
\centering
\begin{tabular}{l|c|cc|cc|cc}
\toprule
\multirow{2}{*}{Model}           & \multicolumn{1}{c|}{\multirow{2}{*}{Para.}} & \multicolumn{2}{c|}{NumpyEval}                            & \multicolumn{2}{c|}{PandasEval}                                     & \multicolumn{2}{c}{TorchDataEval}                        \\ \cline{3-8} 
                                 & \multicolumn{1}{c|}{}                       & \multicolumn{1}{l}{\tiny pass@1} & \multicolumn{1}{l|}{\tiny pass@10} & \multicolumn{1}{l}{\tiny pass@1} & \multicolumn{1}{l|}{\tiny pass@10}           & \multicolumn{1}{l}{\tiny pass@1} & \multicolumn{1}{l}{\tiny pass@10} \\ 
\midrule
\textit{\textbf{General Models}}       &                       & \multicolumn{1}{l}{}       & \multicolumn{1}{l|}{}        & \multicolumn{1}{l}{}       & \multicolumn{1}{l|}{}                  & \multicolumn{1}{l}{}       & \multicolumn{1}{l}{}        \\
CodeT5                           & 220M                  & 0                          & 0.1                          & 0                          & 0                                      & 0                          & 0                           \\
PyCodeGPT                        & 110M                  & 18.04                      & 38.61                        & 12.75                      & 37.62                                   & 3.80                        & 14.00                          \\
CodeGen350M                      & 350M                  & 18.51                & 43.56                  & 16.73                & 29.70                             & 4.60                        & 14.00                          \\
CodeGen2B                        & 2B                    & 29.10                & 53.46                  & 30.69                & 42.57                            & 7.00                          & 18.00                          \\
GPT3.5                           & -                     & 58.41                & 66.21                  & 30.09                 & 33.16                            & 6.00                          & 24.00                          \\ 
\midrule
\textit{\textbf{API-oriented}}    &                       & \multicolumn{1}{l}{}       & \multicolumn{1}{l|}{}        & \multicolumn{1}{l}{}       & \multicolumn{1}{l|}{}                  & \multicolumn{1}{l}{}       & \multicolumn{1}{l}{}        \\
CERT-numpy & 220M             & 31.47                      & 46.42                        & 16.03                & 27.72                            & 2.20                        & 14.00                          \\
CERT-pandas & 220M             & 18.81               & 33.66                  & 28.42                      & 48.04                                  & 2.80                        & 6.00                           \\
CodeGenAPI                   & 350M                  & 16.55                      & 29.48                        & 13.58                      & 34.95                                  & 7.19                       & 16.93                       \\
CodeGenAPI-retrieval         & 475M             & 12.67                      & 27.32                        & 11.25                      & 28.61                                  & 10.41                      & 23.50                        \\
CodeGen-retrieval            & 475M             & 18.30                       & 35.12                        & 9.54                       & 29.02                                  & 7.52                       & 16.36                       \\ 
\midrule
\textit{\textbf{Ours}}           &                       & \multicolumn{1}{l}{}       & \multicolumn{1}{l|}{}        & \multicolumn{1}{l}{}       & \multicolumn{1}{l|}{}                  & \multicolumn{1}{l}{}       & \multicolumn{1}{l}{}        \\
\multirow{2}{*}{ToolCoder-OnlineTool}    & 350M                  & 35.64                & 50.50                  & 22.77                & 37.62                            & 7.40                        & 20.00                          \\
      & 2B                  & 41.58                & 55.44                  & 31.68                & 47.52                            & 11.80                       & 24.00               \\
\bottomrule

\end{tabular}

   \label{tab:rst_public}
\end{table}

To answer RQ1, we evaluate baselines and our ToolCoder on \textit{NumpyEval}, \textit{PandasEval} and \textit{TorchDataEval} and results are shown in Table \ref{tab:rst_public}.
\textit{ToolCoder-OnlineTool} represents the performance of our model with the online search engine tool to generate code.

We notice that some general code generation models, such as CodeT5, have achieved poor results, which proves that the selection of public library API has particular challenges for code generation models. Results show that ToolCoder achieves the best results among general code generation baselines and API-oriented baselines. Even compared with the extremely large model GPT3.5, our model can achieve comparable performance with these public library benchmarks. 

Compared with the state-of-the-art API-oriented baselines, our model achieves 10.11\%, 3.26\%, and 1.39\% pass@1 improvement over the best baseline on three benchmarks. Even when we control our model parameters to be smaller than the baselines as ToolCoder-350M, our model can still achieve excellent overall performances. Existing API-oriented models mainly focus on training and inference on a library API code dataset, resulting in the failure of the same model to achieve good results on multiple API benchmarks, such as CERT-numpy and CERT-pandas. Our model shows stronger generalization ability and can be applied to various API libraries. Our method can achieve excellent results even on the unseen TorchData library.
Our model is trained based on CodeGen models. The performance of our ToolCoder models is significantly higher than that of the corresponding base CodeGen model, indicating that our training process and tool assistant can help models learn to generate API-related code better.

\subsection{RQ2: Results for Private library API Code Generation}
\label{sec:private_rst}
\begin{table}[t]
\scriptsize
    \setlength\tabcolsep{4pt}
    \centering
    \caption{Pass rate of models on Private library benchmarks}
\centering
\begin{tabular}{l|c|cc|cc}
\toprule
\multirow{2}{*}{Model}             & \multirow{2}{*}{Para.} & \multicolumn{2}{c|}{MonkeyEval}                           & \multicolumn{2}{c}{BeatNumEval}                          \\ \cline{3-6} 
                                   &                        & \multicolumn{1}{l}{\tiny pass@1} & \multicolumn{1}{l|}{\tiny pass@10} & \multicolumn{1}{l}{\tiny pass@1} & \multicolumn{1}{l}{\tiny pass@10} \\ \hline
\textit{\textbf{General Models}}   &                        & \multicolumn{1}{l}{}       & \multicolumn{1}{l|}{}        & \multicolumn{1}{l}{}       & \multicolumn{1}{l}{}        \\
CodeT5                             & 220M                   & 0                          & 0                            & 0                          & 0                           \\
CodeGen350M                        & 350M                   & 0.95                       & 4.90                          & 5.15                       & 11.96                       \\
CodeGen2B                          & 2B                     & 1.59                          & 5.94                            & 5.94                          & 11.88                           \\
GPT3.5                             & -                      & 2.47                         & 8.91                            & 6.68                          & 17.82                           \\ \midrule
\textit{\textbf{API-oriented}}     &                        & \multicolumn{1}{l}{}       & \multicolumn{1}{l|}{}        & \multicolumn{1}{l}{}       & \multicolumn{1}{l}{}        \\
CodeGenAPI                     & 350M                   & 1.19                       & 4.68                         & 4.44                       & 8.24                        \\
CodeGenAPI-retrieval           & 475M                   & 3.41                       & 8.33                         & 5.90                        & 11.79                       \\
CodeGen-retrieval              & 475M                   & 2.46                       & 6.35                         & 6.65                       & 13.68                       \\ \midrule
\textit{\textbf{Ours}}             &                        & \multicolumn{1}{l}{}       & \multicolumn{1}{l|}{}        & \multicolumn{1}{l}{}       & \multicolumn{1}{l}{}        \\
\multirow{2}{*}{ToolCoder-DocTool} & 350M                   & 2.98                          & 5.94                            & 6.73                          & 12.87                           \\
                                   & 2B                     & 3.02                          & 7.92                            & 6.93                          & 13.86      \\
                                   \bottomrule
\end{tabular}

   \label{tab:rst_private}
\end{table}

To answer RQ2, we evaluate baselines and our ToolCoder on MonkeyEval and BeatNumEval. Results are shown in Table \ref{tab:rst_private}.
\textit{ToolCoder-DocTool} represents the performance of our model with the documentation search tool to generate code as these private do not have relevant online resources.

These private library benchmarks are extremely hard for general code generation models, which we can see by the smaller pass@1 and pass@10 scores. With the documentation search tool enhanced, our ToolCoder shows stable generalization ability on these two new benchmarks. When compared with the state-of-the-art API-oriented baselines, our model shows comparable performance. Combining the excellent performance of our method on the public library benchmarks, the average \textit{pass@1} on five benchmarks of our two series of ToolCoder is 15.10\%, 19.00\%. For this average pass@1 metric, our ToolCoder outperforms the best baseline CodeGen-retrieval, which is only 8.89\%, raising at least 6.21\% improvement. As for the average \textit{pass@10}, our model outperforms all API-oriented baselines by at least 9.64\%. It is confident that our ToolCoder shows the overall best performance on various API selection scenarios. 

Compared with the base pre-trained model CodeGen-350M and CodeGen-2B, our model greatly improves. ToolCoder-350M outperforms the base CodeGen-350M by 2.03\%, 1.58\% on pass@1 and 1.04\%, 0.91\% on pass@10. ToolCoder-2B also achieves a similar improvement compared with CodeGen-2B. It shows that documentation search tools can help code generation models select proper APIs during inference, thus improving the quality of the generated code. Compared with the most powerful model GPT3.5, our ToolCoder can still achieve better results in some inference settings.
Results show that our proposed ToolCoder can assist the API selection process and enhance the ability of the code generation model.

\subsection{RQ3: Ablation Studies}
\label{sec:ablation}
To answer RQ3, we investigate the impact of different designed modules in our pipeline. We conduct ablation studies, including changing the dataset, training, and inference settings in our experiments.

\subsubsection{Dataset Setting}
\label{sec:ablation_dataset}
\begin{table}[t]
\scriptsize
    \setlength\tabcolsep{5pt}
    \centering
    \caption{Ablation studies on Dataset Settings. We conduct experiments on ToolCoder-350M.}
\centering

\begin{tabular}{l|rr|rr|rr}
\toprule
                                                                            & \multicolumn{2}{c|}{NumpyEval}                            & \multicolumn{2}{c|}{PandasEval}                           & \multicolumn{2}{c}{TorchDataEval}                        \\ \cline{2-7} 
\multirow{-2}{*}{Dataset Setting}                                           & \multicolumn{1}{l}{\tiny pass@1} & \multicolumn{1}{l|}{\tiny pass@10} & \multicolumn{1}{l}{\tiny pass@1} & \multicolumn{1}{l|}{\tiny pass@10} & \multicolumn{1}{l}{\tiny pass@1} & \multicolumn{1}{l}{\tiny pass@10} \\ 
\midrule
ToolCoder-350M                                                          & 35.64                      & 50.50                        & 22.77                      & 37.62                        & 7.40                        & 20.00                          \\
original dataset                                                            & 19.40                          & 39.60                            & 19.92                          & 38.61                            & 6.00                          & 14.00                           \\
annotation w/o query & 14.05                      & 43.56                        & 11.68                      & 33.66                        & 3.80 &  6.00   \\
\midrule
CodeGen-350M                                  & 18.51                & 43.56                  & 16.73                & 29.70                             & 4.60                        & 14.00                          \\
\bottomrule
\end{tabular}
   \label{tab:rst_ablation_dataset}
\end{table}

We perform ablation experiments on the dataset construction in Table \ref{tab:rst_ablation_dataset}. We replace our training dataset with the original dataset, which only contains the regular source code and without annotation, referring as \textit{original dataset}. We also add an experiment to remove the content of the query in the search call so that its form becomes \textit{APISearch()$\rightarrow$answer}. During inference, we use the question description to search the API directly. We refer to this ablation as \textit{annotation w/o query}. We also add the original \textit{CodeGen-350M} model for comparison, which is not trained on the new dataset.

Results show that our dataset annotation is essential for improvement. Compared with the model trained on the original dataset, our ToolCoder-350M shows a stable improvement on almost all metrics. The annotation dataset enables our model to use the external search tool for API selection and thus improve the quality of the generated code. Results also show that it is essential to generate the search query. When we discard the search query in the data construction and use the problem description for API search tools, we observe a drastic drop in the final results as \textit{annotation w/o query} in the Table \ref{tab:rst_ablation_dataset}. We attribute it to the fact that the problem description is still far from the use of the specific API, so it is still difficult to select the appropriate API using the existing API search tools. We can also confirm that only fine-tuning on the original source code dataset can not help the model learn to select APIs. We compare the \textit{CodeGen-350M} with the model trained on the \textit{original dataset}. Results show that additional training on the code dataset does not significantly improve the model's performance. The key to our improvement is to annotate the API tool into the code dataset to teach the model to use external API search tools.

\subsubsection{Training Setting}
\label{sec:ablation_training}
\begin{table}[t]
\scriptsize
    \setlength\tabcolsep{2pt}
    \centering
    \caption{Ablation studies on Training Settings. We conduct experiments on ToolCoder-350M.}
\centering

\begin{tabular}{l|c|c|cc|cc|cc}
\toprule
\multirow{2}{*}{Training Setting} &\multirow{2}{*}{\begin{tabular}[c]{@{}c@{}}Training\\ Time\end{tabular}}  & \multirow{2}{*}{\begin{tabular}[c]{@{}c@{}}Training\\ Para.\end{tabular}}& \multicolumn{2}{c|}{NumpyEval}                            & \multicolumn{2}{c|}{PandasEval}                           & \multicolumn{2}{c}{TorchDataEval}                        \\ \cline{4-9} 
                    &     &         & \multicolumn{1}{l}{\tiny pass@1} & \multicolumn{1}{l|}{\tiny pass@10} & \multicolumn{1}{l}{\tiny pass@1} & \multicolumn{1}{l|}{\tiny pass@10} & \multicolumn{1}{l}{\tiny pass@1} & \multicolumn{1}{l}{\tiny pass@10} \\ 
\midrule
ToolCoder-350M               &    6h  &  0.65M & 35.64                      & 50.50                        & 22.77                      & 37.62                        & 7.40                        & 20.00                          \\
full-training         &  29h   &   350M    & 35.35                      & 58.41                        & 22.67                      & 40.59                        & 6.00                          & 22.00         \\
\bottomrule
\end{tabular}

   \label{tab:rst_ablation_training}
\end{table}

We performed ablation experiments with ToolCoder-350M on the training setting in Table \ref{tab:rst_ablation_training}. Our experiments compare the performance of two approaches: full parameter training, referred to as \textit{full-training}. Our proposed method utilizes LoRA for parameter-efficient training. We evaluate their performance on public library benchmarks and recorded their training costs, including training time and parameters, using 2*2080 GPUs.

Results show that our fine-tuning strategy has almost no performance penalty compared with the regular \textit{full-training}. On the public library benchmarks, the difference between the two pass@1 results is within 0.4\%. The gap in these results is acceptable, considering the huge savings in training costs. In our experiment settings, our parameter-efficient fine-tuning strategy can reduce the training time from 29h to 6h and the training parameters from more than 350M to 0.65M. We only need to train 0.18\% parameters in CodeGen-350M and 0.09\% for CodeGen-2B. It makes it possible to efficiently fine-tune models on a consumer-level GPU, such as Nvidia GeForce RTX 2080 (11GB RAM).

\subsubsection{Inference Setting}
\label{sec:ablation_infer}
\begin{table}[t]
    \centering
    \caption{Ablation studies on Inference Settings.}
\centering

\subtable[{On Public library benchmarks}]{
\centering
\scriptsize
\begin{tabular}{l|cc|cc|cc}
\toprule
\multirow{2}{*}{Inference Setting}  & \multicolumn{2}{l|}{NumpyEval} & \multicolumn{2}{l|}{PandasEval} & \multicolumn{2}{l}{TorchDataEval} \\ \cline{2-7} 
                                            & {\tiny pass@1}       & {\tiny pass@10}        & {\tiny pass@1}       & {\tiny pass@10}        & {\tiny pass@1}       & {\tiny pass@10}         \\ \midrule
OnlineTool-350M                            & 35.64         & 50.50          & 22.77          & 37.62          & 7.40             & 20.00              \\
NoTool-350M                                    & 33.76         & 46.53          & 20.19          & 35.64          & 6.00               & 16.00              \\ \midrule
OnlineTool-2B                               & 41.58         & 55.44          & 31.68          & 47.52          & 11.80            & 24.00 \\
NoTool-2B                                   & 38.71         & 54.45          & 31.38          & 44.55          & 7.50             & 20.00         \\
\bottomrule
\end{tabular}
}
\subtable[{On Private library benchmarks}]{
\centering
\scriptsize
\begin{tabular}{l|cc|cc}
\toprule
\multirow{2}{*}{Inference Setting}  & \multicolumn{2}{l|}{MonkeyEval} & \multicolumn{2}{l}{BeatNumEval}  \\ \cline{2-5} 
                                     & {\tiny pass@1}       & {\tiny pass@10}        & {\tiny pass@1}       & {\tiny pass@10}                 \\ \midrule
OnlineTool-350M                                 & 2.98         & 5.94          & 6.73          & 12.87                        \\
NoTool-350M                                    & 0.29         & 0.99          & 1.68          & 4.95                        \\ \midrule
OnlineTool-2B                                 & 3.02                          & 7.92                            & 6.93                          & 13.86                      \\
NoTool-2B                                     & 0.79         & 2.97          & 2.77          & 8.91               \\
\bottomrule
\end{tabular}
}
   \label{tab:rst_ablation_infer}
\end{table}

We perform ablation experiments on the inference setting in Table \ref{tab:rst_ablation_infer}. We add experiments to disable the tool in our model. \textit{NoTool} represents that we disable the tool for inference and use our trained model to directly generate an API based on the search query and complete the code. We compare with our original inference setting on public and private library benchmarks. 

Experiments show that our external tools are essential in improving performance. On public library benchmarks, the online search engine tool improves pass@1 by 1.88\%, 2.57\%, 0.4\% for ToolCoder-350M, and 2.87\%, 0.29\%, 4.3\% for ToolCoder-2B. The online search engine tool can search for similar API usage scenarios and provide accurate API suggestions.
When considering private library benchmarks, the improvement is more significant. We find the model itself works poorly on private libraries. However, with the assistance of the documentation search tool, our model can choose a suitable private library API.

Another interesting observation is that the \textit{NoTool} also achieves relatively good performance on public library benchmarks. We believe that the improvement comes from our dataset annotation process. The additional tool call process in the dataset can be seen as a way to think about and choose the API. The chain of thought in the annotation dataset can assist the code generation model in better understanding the functions and application scenarios of different APIs, thus directly improving the model to select the API. However, for private libraries, since the knowledge of private libraries is not seen by the code generation model, this form of dataset annotation is challenging to bring improvements to the model. With proper API search tools enhanced, our ToolCoder can select API more accurately and improve further.

\subsection{RQ4: Qualitative analysis}
\begin{figure}[t]
\centering
  \includegraphics[width=0.7\columnwidth]{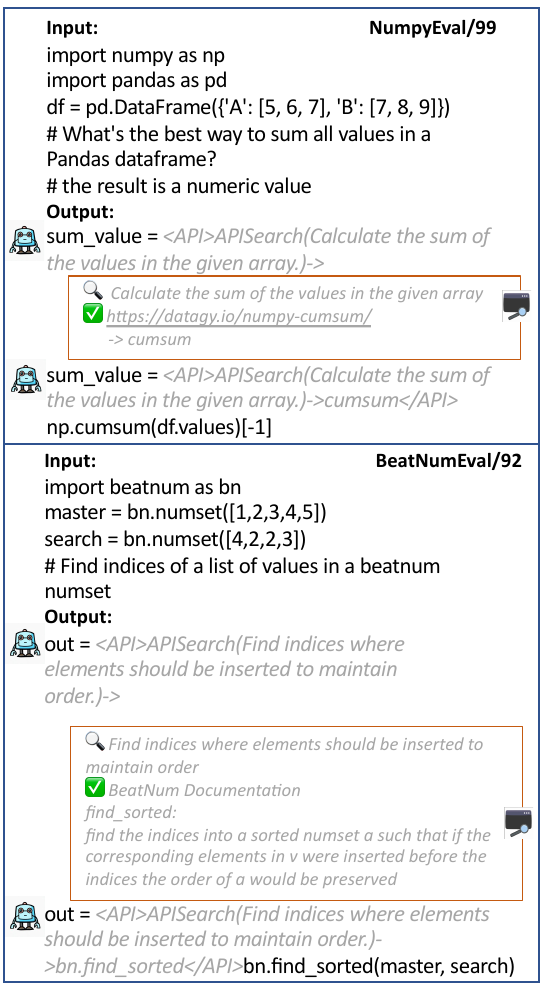}  
\caption{Case Studies of the ToolCoder-2B, with online search engine tool on NumpyEval and documentation search tool on BeatNumEval.}
\label{fig:case_study}
\end{figure}
To answer RQ4, we perform a case study analysis to assess the generated code’s quality. 
Figure \ref{fig:case_study} represents code snippets generated on public and private library benchmarks. From the examples, we obtain the following findings: \ding{182} The generated search query provides more fine-grained technical planning for the solution. The \textit{NumpyEval} case requires summing values in a dataframe, and the generated query breaks down the requirements, focusing first on summing arrays. It fills the gap between requirements and concrete APIs. \ding{183} The response of the search tools both play a crucial role in the generated code. The online search engine tool finds the proper API from the correct websites, and the documentation search tool finds the proper API by searching over the API comments. \ding{184} Our ToolCoder also can make necessary modifications based on the tool response. For example, the online search tool returns the response as \textit{cumsum}, not directly defined in the input code. Our ToolCoder can add some components not in the response and generate the correct API \textit{np.cumsum}.

\section{Threats to Validity}
\label{sec:validity}
\paragraphwithoutdot{Threats to internal validity} are related to the roles of the model architecture and hyper-parameters setting.
In our experiments, we do a small-range grid search on learning rate and batch size settings. Our ToolCoder-350M model tries to keep the hyper-parameters the same as baseline models for a fair comparison.

\paragraphwithoutdot{Threats to external validity} are mainly related to the tasks and datasets we choose in this paper. We counter this by evaluating our model on five different benchmarks of two types of API, including public and private library API code generation.

\paragraphwithoutdot{Threats to construct validity} include the evaluation metrics we used in this work. We utilize pass rates to evaluate the correctness of generated code accurately. This metric is adequate for corresponding tasks and has been adopted by many previous studies.

\section{Related Work}
\label{sec:related}
\subsection{Code Generation}
Code generation aims to generate the source code that satisfies a given natural language description or requirement.
It involves automatically creating source code based on functional requirements, such as natural language descriptions \cite{DBLP:conf/emnlp/IyerKCZ18} or pseudo code algorithms \cite{kulal2019spoc, oda2015django, yin2018tranx}. 
Recently pre-trained language models have shown impressive capabilities in code generation tasks. 
Lu et al. \cite{codexglue} adapt GPT-2 \cite{gpt2} model on the source code, resulting in CodeGPT. Chen et al. \cite{humaneval} fine-tune GPT-3 \cite{gpt3} models on the code to produce CodeX and GitHub Copilot. OpenAI also produces the GPT3.5 series of models, which have shown strong generation capabilities in natural language and programming languages. Neither CodeX nor GPT3.5 is open-sourced, which leads to several attempts to replicate CodeX in industry and academia, resulting in GPT-Neo \cite{gpt-neo}, GPT-J \cite{gpt-j}, CodeParrot \cite{wolf-etal-2020-transformers}, PolyCoder \cite{PolyCoder}, PyCodeGPT \cite{cert}, InCoder \cite{incoder}, and CodeGen \cite{codegen}. In our experiments, we choose the CodeGen series of models as our base model for further exploration.

Recently, some work has focused on selecting APIs during code generation. As discussed in Section \ref{sec:motivating_codegen}, existing code generation models still struggle with selecting appropriate APIs for a given context, especially for private or lesser-known APIs. Existing work \cite{cert, codegenapi, zhou2023docprompting} has proposed some API-oriented code generation methods. They typically use a two-stage pipeline, where the first stage involves searching or generating related APIs and then using them to generate code.
We pursue this research line and propose to leverage pre-trained models and API search tools to automate API selection in coding practices. 
In comparison, our approach has two advantages: \ding{182} Our method shows strong generalization ability. By setting an appropriate API search tool, our method can quickly adapt to any API-related code generation scenario. \ding{183} Our method does not require multi-stage generation. Instead, we integrate the API search tool into the decoding process, making our approach more flexible and allowing the API selection process to be closer to the specific code fragment being generated.

\subsection{Tool-Augmented Large Language Models}
Recent research in language modeling has explored using external tools to supplement the knowledge stored in the model’s weights \cite{DBLP:journals/corr/abs-2302-07842}. These external tools can include other neural networks or even the language model itself, allowing for the composition of different pre-trained models on various modalities, such as the Socratic Model \cite{DBLP:journals/corr/abs-2204-00598}.
Alternatively, natural language knowledge can be retrieved from external sources, as demonstrated by WebGPT \cite{webgpt} and ReAct \cite{react} through the use of search APIs.
Other approaches, such as Toolformer \cite{toolformer} and ART \cite{art}, leverage a combination of search tools, question-answering tools, machine translation tools, calculators, and other tools to solve various NLP tasks.
ChatGPT Plugins\footnote{https://openai.com/blog/chatgpt-plugins} further demonstrate the potential for language models to integrate with thousands to millions of tools.
However, incorporating programming tools into code-related models has not been explored yet. Our paper addresses this gap by abstracting the process of human programmers selecting APIs into a programming tool that augments code generation models.

\section{Conclusion}
\label{sec:conclusion}
In this paper, we propose ToolCoder, a novel approach incorporating API search tools into the code generation process to assist models in selecting appropriate APIs. 
We categorize API search tools into two types, including online search engine tools and documentation search tools, and abstract them into a unified form. 
We propose an automatic dataset annotation method to add tool usage information to the source code data. 
The parameter-efficient strategy is used to fine-tune the model. 
During inference, the model decoding process is enhanced with external API search tools for proper API selection. 
Experiments on public and private library code generation benchmarks show that our ToolCoder outperforms state-of-the-art methods, with at least a 6.21\% improvement on average pass@1 metrics. 
Our experiments also demonstrate the potential of incorporating programming tools into the code generation process, shedding light on this line of future work.

\bibliographystyle{ACM-Reference-Format}
\bibliography{sample-base}

\end{document}